# New development in theory of Laguerre polynomials

I. I. Guseinov

*Department of Physics, Faculty of Arts and Sciences, Onsekiz Mart University, Çanakkale, Turkey*

**Abstract**

The new complete orthonormal sets of $\mathcal{L}^\alpha$-Laguerre type polynomials ($\mathcal{L}^\alpha$-LTP, $\alpha = 2, 1, 0, -1, -2,...$) are suggested. Using Schrödinger equation for complete orthonormal sets of $\psi^\alpha$-exponential type orbitals ($\psi^\alpha$-ETO) introduced by the author, it is shown that the origin of these polynomials is the centrally symmetric potential which contains the core attraction potential and the quantum frictional potential of the field produced by the particle itself. The quantum frictional forces are the analog of radiation damping or frictional forces suggested by Lorentz in classical electrodynamics. The new $\mathcal{L}^\alpha$-LTP are complete without the inclusion of the continuum states of hydrogen like atoms. It is shown that the nonstandard and standard conventions of $\mathcal{L}^\alpha$-LTP and their weight functions are the same. As an application, the sets of infinite expansion formulas in terms of $\mathcal{L}^\alpha$-LTP and L-Generalized Laguerre polynomials (L-GLP) for atomic nuclear attraction integrals of Slater type orbitals (STO) and Coulomb-Yukawa like correlated interaction potentials (CIP) with integer and noninteger indices are obtained. The arrange and rearranged power series of a general power function are also investigated. The convergence of these series is tested by calculating concrete cases for arbitrary values of parameters of orbitals and power function.

**Key Words:** Centrally symmetric potentials, $\psi^\alpha$- exponential type orbitals, Frictional quantum numbers, Generalized Laguerre polynomials

## 1. Introduction

It is well known that the eigenfunctions of the Schrödinger equation for the hydrogen-like atoms the radial parts of which contain the L-GLP are not complete unless the continuum is included [1]. Because of this, some difficulties arise in the solution of different physical problems when the L-GLP are employed (see Ref. [2] and references quoted therein). Therefore, the necessity arises for the construction of the new $\mathcal{L}^\alpha$-LTP using complete orthonormal sets of radial parts of $\psi^\alpha$-ETO [3]. As has been shown in a previous paper [4], the eigenfunctions $\psi^\alpha$-ETO correspond to the total centrally symmetric potential which contains the core attraction potential and the Lorentz frictional potential of the field produced by the particle itself. We notice that the Lambda and Coulomb-Sturmian ETO introduced in Refs. [5-8] are the special cases of the $\psi^\alpha$-ETO for $\alpha = 0$ and $\alpha = 1$, respectively.

## 2. Definition and basic formulas

The new $\mathcal{L}^\alpha$-LTP in non-standard conventions suggested in this work are defined as

$$\mathcal{L}_{nl}^\alpha(x) = (-1)^\alpha \left[\frac{(q-p)!}{(2n)^\alpha (q!)^3}\right] x^l L_q^p(x) = \sum_{k=l}^{n-1} \Pi_{nk}^{\alpha l} x^k, \tag{1}$$

where

$$\Pi_{nk}^{\alpha l} = (-1)^{k-l} \left[\frac{(q-p)!}{(2n)^\alpha q!}\right]^{1/2} \frac{F_{p+k-l}(q)}{(k-l)!} \tag{2a}$$

$$F_m(n) = \begin{cases} \dfrac{n!}{m!(n-m)!} & for\ 0 \le m \le n \\ 0 & for\ m < 0\ and\ m > n. \end{cases} \tag{2b}$$
$$\tag{2c}$$

Here, $\alpha$ is the frictional quantum number $(-\infty < \alpha \le 2)$ (see Ref. [4]), $p = 2l + 2 - \alpha$, $q = n + l + 1 - \alpha$ and $L_q^p(x)$ are the non-standard L-GLP. The $\mathcal{L}^\alpha$-LTP are orthonormal with respect to the weight function $\left(\dfrac{2n}{x}\right)^\alpha$:

$$\int_0^\infty e^{-x} x^2 \mathcal{L}_{nl}^\alpha(x) \bar{\mathcal{L}}_{n'l}^\alpha(x) dx = \delta_{nn'} \tag{3}$$

$$\bar{\mathcal{L}}_{nl}^\alpha(x) = \left(\frac{2n}{x}\right)^\alpha \mathcal{L}_{nl}^\alpha(x). \tag{4}$$

Now we take into account Eq. (1) in the formulae for radial parts of $\psi^\alpha$-ETO in nonstardard conventions [3] which are the complete orthonormal sets of eigenfunctions of Schrödinger equation for hydrogen-like atoms. Then, we obtain:

$$R_{nl}^\alpha(\zeta, r) = (2\zeta)^{3/2} R_{nl}^\alpha(x) \tag{5a}$$

$$\bar{R}_{nl}^\alpha(\zeta, r) = (2\zeta)^{3/2} \bar{R}_{nl}^\alpha(x), \tag{5b}$$

where $x = 2\zeta r$ and

$$R_{nl}^\alpha(x) = e^{-\frac{x}{2}} \mathcal{L}_{nl}^\alpha(x) \tag{6a}$$

$$\overline{R}_{nl}^{\alpha}(x) = e^{-\frac{x}{2}} \overline{\mathcal{L}}_{nl}^{\alpha}(x). \tag{6b}$$

We note that the similar formulas can also be derived in standard conventions using the relation

$$L_q^p(x) = (-1)^p q! L_{q-p}^{(p)}(x) = (-1)^{\alpha} q! L_{q-p}^{(p)}(x), \tag{7}$$

where $L_{q-p}^{(p)}(x)$ are the standard L-GLP. The non-standard and standard conventions for the L-GLP were discussed in Refs. [9] and [10], respectively (see Ref. [11] for the definition of $\psi^{\alpha}$-ETO in standard conventions). Taking into account Eqs. (1) and (7) we obtain for the $\mathcal{L}^{\alpha}$-LTP in standard conventions the following relations:

$$\mathcal{L}_{nl}^{(\alpha)}(x) = (-1)^{\alpha} q! N_{nl}^{\alpha} x^l L_{q-p}^{(p)}(x) \tag{8a}$$

$$\overline{\mathcal{L}}_{nl}^{(\alpha)}(x) = \left(\frac{2n}{x}\right)^{\alpha} \mathcal{L}_{nl}^{(\alpha)}(x). \tag{8b}$$

It is easy to show that the $\mathcal{L}^{(\alpha)}$ are orthogonal with respect to the weight function $\left(\frac{2n}{x}\right)^{\alpha}$:

$$\int_0^{\infty} e^{-x} x^2 \mathcal{L}_{nl}^{(\alpha)}(x) \overline{\mathcal{L}}_{n'l}^{(\alpha)}(x) dx = \delta_{nn'}. \tag{9}$$

The radial parts of complete ortonormal sets of $\psi^{(\alpha)}$-ETO in standard conventions through the $\mathcal{L}^{(\alpha)}$ are defined as

$$R_{nl}^{(\alpha)}(\zeta, r) = (2\zeta)^{3/2} R_{nl}^{(\alpha)}(x) \tag{10a}$$

$$\overline{R}_{nl}^{(\alpha)}(\zeta, r) = (2\zeta)^{3/2} \overline{R}_{nl}^{(\alpha)}(x), \tag{10b}$$

where

$$R_{nl}^{(\alpha)}(x) = e^{-\frac{x}{2}} \mathcal{L}_{nl}^{(\alpha)}(x) \tag{11a}$$

$$\overline{R}_{nl}^{(\alpha)}(x) = e^{-\frac{x}{2}} \overline{\mathcal{L}}_{nl}^{(\alpha)}(x). \tag{11b}$$

It is easy to show that the nonstandard $\mathcal{L}^{\alpha}$ and standard $\mathcal{L}^{(\alpha)}$ and their weight functions $\overline{\mathcal{L}}^{\alpha}$ and $\overline{\mathcal{L}}^{(\alpha)}$, respectively, are the same, i.e., $\mathcal{L}^{\alpha} = \mathcal{L}^{(\alpha)}$ and $\overline{\mathcal{L}}^{\alpha} = \overline{\mathcal{L}}^{(\alpha)}$.

The nonstandard $\mathcal{L}^\alpha$- and standard $\mathcal{L}^{(\alpha)}$-LTP form the complete orthonormal sets on the interval $(0,\infty)$ with the weight functions $\bar{\mathcal{L}}^\alpha$ and $\bar{\mathcal{L}}^{(\alpha)}$, respectively. The completeness relations can be proved by the use of method set out in Ref. [10] (see also [12]). It is easy to show that

$$e^{-\frac{1}{2}(x+x')}\sum_{n=l+1}^{\infty}\mathcal{L}_{nl}^{\alpha}(x)\bar{\mathcal{L}}_{nl}^{\alpha}(x')=\sum_{n=l+1}^{\infty}R_{nl}^{\alpha}(x)\bar{R}_{nl}^{\alpha}(x')=\delta(x-x') \qquad (12)$$

$$e^{-\frac{1}{2}(x+x')}\sum_{n=l+1}^{\infty}\mathcal{L}_{nl}^{(\alpha)}(x)\bar{\mathcal{L}}_{nl}^{(\alpha)}(x')=\sum_{n=l+1}^{\infty}R_{nl}^{(\alpha)}(x)\bar{R}_{nl}^{(\alpha)}(x')=\delta(x-x'). \qquad (13)$$

It should be noted that, in the special cases of the $\mathcal{L}^\alpha$- and $\mathcal{L}^{(\alpha)}$-LTP for $\alpha=0$ and $\alpha=1$, the Eqs. (12) and (13) describe the completeness properties of Lambda and Coulomb-Sturmian functions with nonstandard and standard conventions, respectively (see Refs.[5-8]).

## 3. Differential equation of $\mathcal{L}^\alpha$-LTP

In order to derive the differential equation for $\mathcal{L}^\alpha$-LTP we use the Schrödinger equation for the radial parts $R_{nl}^\alpha(x)$ in the form

$$2\zeta^2\left[-\frac{1}{x^2}\frac{d}{dx}\left(x^2\frac{d}{dx}\right)+\frac{l(l+1)}{x^2}+\frac{1}{2\zeta^2}V(x)\right]R_{nl}^\alpha(x)=\varepsilon R_{nl}^\alpha(x), \qquad (14)$$

where $\varepsilon=-\zeta^2/2$ (see Ref. [4]). Here, $V(\zeta,r)$ denotes the potential of the centrally symmetric field which corresponds to the eigenfunctions $\psi^\alpha$-ETO. The substitution of Eq. (6a) into (14) yields the following equation for the $\mathcal{L}^\alpha$-LTP:

$$x\frac{d^2\mathcal{L}_{nl}^\alpha}{dx^2}+(3-\alpha-x)\frac{d\mathcal{L}_{nl}^\alpha}{dx}-(1-\alpha)\frac{d\mathcal{L}_{nl}^\alpha}{dx}-\left(1+\frac{l(l+1)}{x}+\frac{x}{2\zeta^2}V(x)\right)\mathcal{L}_{nl}^\alpha=0. \qquad (15)$$

Next we use the formula (1) in the equation

$$x\frac{d^2L_q^p}{dx^2}+(p+1-x)\frac{dL_q^p}{dx}+(q-p)L_q^p=0 \qquad (16)$$

for non-standard L-GLP [13] and the condition

$$\frac{d^k L_q^p}{dx^k}=L_q^{p+k}. \qquad (17)$$

Then, a simple algebra leads to

$$x\frac{d^2\mathcal{L}_{nl}^\alpha}{dx^2}+(3-\alpha-x)\frac{d\mathcal{L}_{nl}^\alpha}{dx}-\left((1-n)+\frac{l(l+1)}{x}+\frac{l(1-\alpha)}{x}\right)\mathcal{L}_{nl}^\alpha=0. \qquad (18)$$

Thus, we obtained for the $\mathcal{L}^\alpha$-LTP two kinds of independent equations one of which, Eq. (15), contains the potential $V(x) \equiv V_{nl}^\alpha(\zeta,r)$. The comparison of these equations gives

$$V_{nl}^\alpha(\zeta,r) = U_n(\zeta,r) + U_{nl}^\alpha(\zeta,r), \tag{19}$$

where the first and second terms are the core attraction and frictional potentials, respectively,

$$U_n(\zeta,r) = -\frac{2\zeta^2}{x}n \tag{20}$$

$$U_{nl}^\alpha(\zeta,r) = -\frac{2\zeta^2(1-\alpha)}{x}\left(\frac{d\mathcal{L}_{nl}^\alpha(x)}{dx}\bigg/\mathcal{L}_{nl}^\alpha(x) - \frac{l}{x}\right) \tag{21a}$$

$$= -\frac{2\zeta^2(1-\alpha)}{x}L_q^{p+1}(x)\big/L_q^p(x), \tag{21b}$$

where $x = 2\zeta r$.

Here, the function $U_{nl}^\alpha$ is the Lorentz frictional self-potential of the field produced by the particle at the point where it is located. See Ref. [4] for the description of properties of these potentials.

Thus, we have established a large number ($\alpha = 2,1,0,-1,-2,...$) of independent complete orthonormal sets of relations for $\mathcal{L}^\alpha$-LTP the origin of which is the core attraction and frictional potentials of the field produced by the particle itself.

## 4. Use of $\mathcal{L}^\alpha$-LTP and L-GLP in study of power series of a general power function

In this section we consider the arrange and rearranged power series of a function

$$f^{\mu^*}(\xi,r) = r^{\mu^*-1}e^{-\xi r} = r^{n+\eta^*}e^{-\xi r}, \tag{22}$$

where $n$ is the integer part of $\mu^*-1$ and $0<\eta^*<1$. For this purpose we utilize the following Laguerre series obtained with the help of $\mathcal{L}^\alpha$-LTP and L-GLP:

for $\mathcal{L}^\alpha$-LTP

$$r^{\eta^*}e^{-\xi r} = \sum_{\mu=\nu+1}^\infty A_{\eta^*\mu}^{\alpha\nu}(\xi)\mathcal{L}_{\mu\nu}^\alpha(r), \tag{23}$$

for L-GLP

$$r^{\eta^*}e^{-\xi r} = \sum_{\mu=\nu}^\infty B_{\eta^*\mu}^\nu(\xi)L_\mu^\nu(r), \tag{24}$$

where $0 \leq \xi < \infty$, $\alpha = 2,1,0,-1,-2,...$, $\nu = 0,1,2,...$ and

$$A_{\eta^*\mu}^{\alpha\nu}(\xi) = (2\mu)^{\alpha} \sum_{s=\nu}^{\mu-1} \Pi_{\mu s}^{\alpha\nu} \frac{\Gamma(\eta^* - \alpha + s + 3)}{(1+\xi)^{\eta^* - \alpha + s + 3}} \tag{25}$$

$$B_{\eta^*\mu}^{\nu}(\xi) = \frac{(\mu-\nu!)}{(\mu!)^3} \sum_{s=0}^{\mu-\nu} \beta_{\mu s}^{\nu} \frac{\Gamma(\eta^* + \nu + s + 1)}{(1+\xi)^{\eta^* + \nu + s + 1}} \ . \tag{26}$$

Here, $\beta_{\mu s}^{\nu}$ are the Laguerre coefficients determined by

$$L_{\mu}^{\nu}(x) = \sum_{s=0}^{\mu-\nu} \beta_{\mu s}^{\nu} x^s \tag{27}$$

$$\beta_{\mu s}^{\nu} = (-1)^{\nu+s} (\mu-s)! F_s(\mu) F_{\nu+s}(\mu). \tag{28}$$

Now we obtain the power series. For this purpose, we use the method set out in previous papers [3,14]. Then, taking into account the properties (2b) and (2c) in Eqs. (23) and (27) we obtain for the arrange, Eqs. (29a) and (30a), and rearranged, Eqs. (29b) and (30b), power series the following formulae:

for $\mathcal{L}^{\alpha}$-LTP

$$r^{\eta^*} e^{-\xi r} = \sum_{\mu=\nu+1}^{\infty} A_{\eta^*\mu}^{\alpha\nu}(\xi) \sum_{s=\nu}^{\mu-1} \Pi_{\mu s}^{\alpha\nu} r^s = \lim_{N \to \infty} \sum_{\mu=\nu+1}^{N} A_{\eta^*\mu}^{\alpha\nu}(\xi) \sum_{s=\nu}^{\mu-1} \Pi_{\mu s}^{\alpha\nu} r^s \tag{29a}$$

$$= \lim_{N \to \infty} \sum_{\mu=\nu}^{N-1} Q_{\eta^*\mu}^{\alpha\nu}(N,\xi) r^{\mu}, \tag{29b}$$

for L-GLP

$$r^{\eta^*} e^{-\xi r} = \sum_{\mu=\nu}^{\infty} B_{\eta^*\mu}^{\nu}(\xi) \sum_{s=0}^{\mu-\nu} \beta_{\mu s}^{\nu} r^s = \lim_{N \to \infty} \sum_{\mu=\nu}^{N} B_{\eta^*\mu}^{\nu}(\xi) \sum_{s=0}^{\mu-\nu} \beta_{\mu s}^{\nu} r^s \tag{30a}$$

$$= \lim_{N \to \infty} \sum_{\mu=0}^{N-\nu} D_{\eta^*\mu}^{\nu}(N,\xi) r^{\mu}, \tag{30b}$$

where

$$Q_{\eta^*\mu}^{\alpha\nu}(N,\xi) = \sum_{s=\nu+1}^{N} A_{\eta^*s}^{\alpha\nu}(\xi) \Pi_{s\mu}^{\alpha\nu} \tag{31}$$

$$D_{\eta^*\mu}^{\nu}(N,\xi) = \sum_{s=\nu}^{N} B_{\eta^*s}^{\nu}(\xi) \beta_{s\mu}^{\nu} \ . \tag{32}$$

As an example of application of Eqs. (29a), (29b), (30a) and (30b) we calculate the atomic nuclear attraction integrals of STO and Coulomb-Yukawa like CIP with integer and noninteger indices defined by [15]

$$I^{q^*}_{p^*p'^*}(\zeta,\zeta',\xi) = \int \chi^*_{p^*}(\zeta,\vec{r})\chi_{p'^*}(\zeta',\vec{r})f^{q^*}(\xi,\vec{r})d^3\vec{r}, \tag{33}$$

where $p^* \equiv n^*lm$, $p'^* \equiv n'^*l'm'$, $q^* \equiv \mu^*\nu\sigma$ and

$$\chi_{n^*lm}(\zeta,\vec{r}) = R_{n^*}(\zeta,r)S_{lm}(\theta,\varphi) \tag{34}$$

$$R_{n^*}(\zeta,r) = (2\zeta)^{n^*+1/2}\left[\Gamma(2n^*+1)\right]^{-1/2} r^{n^*-1}e^{-\zeta r} \tag{35}$$

$$f^{\mu^*\nu\sigma}(\xi,\vec{r}) = \left(\frac{4\pi}{2\nu+1}\right)^{1/2} f^{\mu^*}(\xi,r)S_{\nu\sigma}(\theta,\varphi) \tag{36}$$

$$f^{\mu^*}(\xi,r) = r^{\mu^*-1}e^{-\xi r}. \tag{37}$$

It is easy to derive for integral (33) the relation

$$I^{q^*}_{p^*p'^*}(\zeta,\zeta',\xi) = C^{\nu|\sigma|}(lm,l'm')A^\sigma_{mm'}I^{\mu^*}_{n^*n'^*}(\zeta,\zeta',\xi), \tag{38}$$

where $C^{\nu|\sigma|}(lm,l'm')$ are the generalized Gaunt coefficients [16] and

$$I^{\mu^*}_{n^*n'^*}(\zeta,\zeta',\xi) = \int_0^\infty R_{n^*}(\zeta,r)R_{n'^*}(\zeta',r)f^{\mu^*}(\xi,r)r^2 dr. \tag{39}$$

These integrals are determined from the following analytical relation:

$$I^{\mu^*}_{n^*n'^*}(\zeta,\zeta',\xi) = N_{n^*n'^*}(\zeta,\zeta')\frac{\Gamma(N^*+\mu^*+1)}{(\varepsilon+\xi)^{N^*+\mu^*+1}}, \tag{40}$$

where $\varepsilon = \zeta+\zeta'$, $N^* = n^*+n'^*-1$ and

$$N_{n^*n'^*}(\zeta,\zeta') = \frac{(2\zeta)^{n^*+1/2}(2\zeta')^{n'^*+1/2}}{\left[\Gamma(2n^*+1)\Gamma(2n'^*+1)\right]^{1/2}}. \tag{41}$$

Now we evaluate the integral (39) using relations (29a), (29b), (30a) and (30b). Then, it is easy to derive for the atomic nuclear attraction integrals the following large number of arrange, Eqs. (42a) and (43a), and rearranged, Eqs. (42b) and (43b), power series expansion relations:

for $\mathcal{L}^\alpha$-LTP

$$I^{\mu^*}_{n^*n'^*}(\zeta,\zeta',\xi) = \sum_{\mu=\nu+1}^\infty A^{\alpha\nu}_{\eta^*\mu}(\xi)\sum_{s=\nu}^{\mu-1}\prod^{\alpha\nu}_{\mu s} J^{n+s}_{n^*n'^*}(\zeta,\zeta') = \lim_{N\to\infty}\sum_{\mu=\nu+1}^N A^{\alpha\nu}_{\eta^*\mu}(\xi)\sum_{s=\nu}^{\mu-1}\prod^{\alpha\nu}_{\mu s} J^{n+s}_{n^*n'^*}(\zeta,\zeta') \tag{42a}$$

$$= \lim_{N\to\infty}\sum_{\mu=\nu}^{N-1} Q^{\alpha\nu}_{\eta^*\mu}(N,\xi)J^{n+\mu}_{n^*n'^*}(\zeta,\zeta'), \tag{42b}$$

for L-GLP

$$I_{n^*n'^*}^{\mu^*}(\zeta,\zeta',\xi) = \sum_{\mu=\nu}^{\infty} B_{\eta^*\mu}^{V}(\xi) \sum_{s=0}^{\mu-\nu} \beta_{\mu s}^{V} J_{n^*n'^*}^{n+s}(\zeta,\zeta') = \lim_{N\to\infty} \sum_{\mu=\nu}^{N} B_{\eta^*\mu}^{V}(\xi) \sum_{s=0}^{\mu-\nu} \beta_{\mu s}^{V} J_{n^*n'^*}^{n+s}(\zeta,\zeta') \quad (43a)$$

$$= \lim_{N\to\infty} \sum_{\mu=0}^{N-\nu} D_{\eta^*\mu}^{V}(N,\xi) J_{n^*n'^*}^{n+\mu}(\zeta,\zeta'), \quad (43b)$$

where $-\infty < \alpha \leq 2$ and $0 \leq \nu < \infty$.

Here, the quantities $J_{n^*n'^*}^{\kappa}(\zeta,\zeta')$ are determined by

$$J_{n^*n'^*}^{\kappa}(\zeta,\zeta') = \int_0^{\infty} R_{n^*}(\zeta,r) R_{n'^*}(\zeta',r) r^{\kappa+2} dr = N_{n^*n'^*}(\zeta,\zeta') \frac{\Gamma(N^* + \kappa + 2)}{\varepsilon^{N^*+\kappa+2}}, \quad (44)$$

where $\kappa = n + s$ and $\kappa = n + \mu$.

## 5. Numerical results and discussion

The applicability of the arrange and rearranged power series obtained from the use of complete orthonormal sets of $\mathcal{L}^{\alpha}$-LTP and L-GLP is tested by calculating the atomic nuclear attraction integrals determined by Eqs. (42a), (42b) and (43a), (43b), respectively. On the basis of these formulae we constructed the programs which are performed in the Mathematica 7.0 language packages. The convergence properties of Coulomb (for $\xi = 0$) and Yukawa (for $\xi = 5.1$) like nuclear attraction integrals for $\nu = 0$ and $-2 \leq \alpha \leq 2$ are shown in Figures 1 and 2, respectively.

The Figures 1 and 2 show a good rate agreement of values obtained from the arrange and rearranged series expansion relations. Thus, the Eqs. (42a), (42b), (43a) and (43b) display the most rapid convergence as a function of summation limits for $N = 40$. We notice that the greater accuracy is attainable by the use of more terms in series expansion relations obtained in this work.

## 6. Conclusion

We have demonstrated that the determination of arrange and rearranged power series of a function $f^{\mu^*}(\xi,r)$ obtained by the use of $\mathcal{L}^{\alpha}$-LTP and L-GLP is legitimate for the integer and noninteger values of indices $\mu^*$. As we see from our tests that the power series derived in this work with the help of $\mathcal{L}^{\alpha}$-LTP and L-GLP can be useful tool for evaluation of the multicenter nuclear attraction integrals when arrange and rearranged one-range addition theorems for STO and Coulomb-Yukawa like CIP presented in our previous papers are employed.

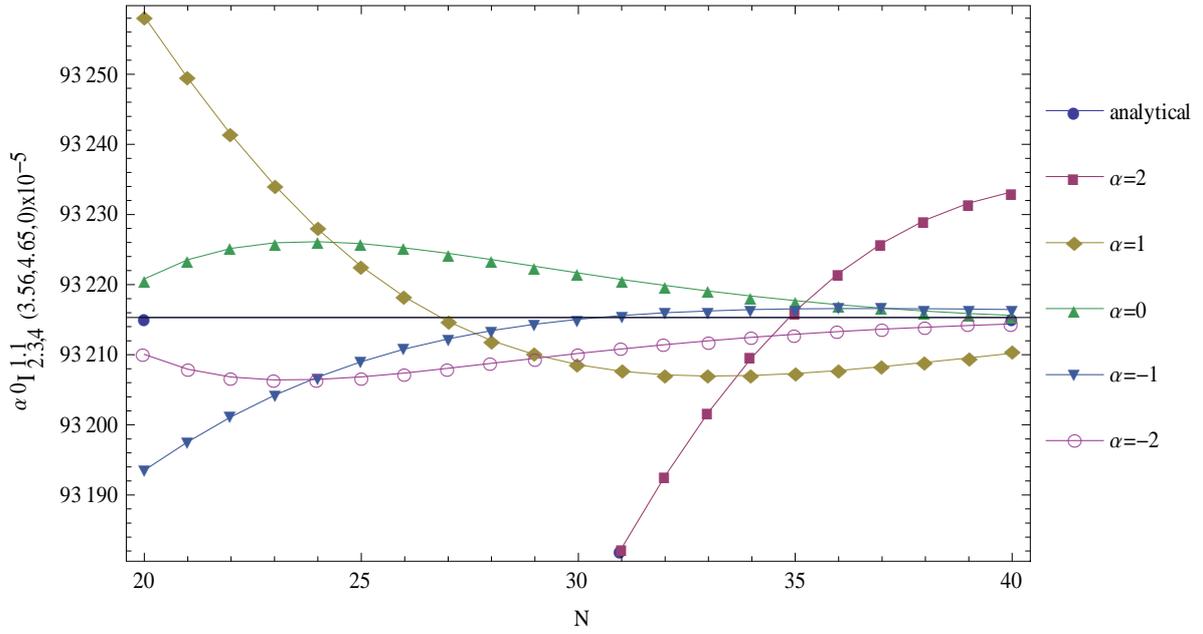

Fig.1. Convergence of arrange (42a) and rearranged (42b) power series for Coulomb like integrals ${}^{\alpha 0}I_{2.3,4}^{1.1}(3.56,4.65,0)$ for $-2 \leq \alpha \leq 2$

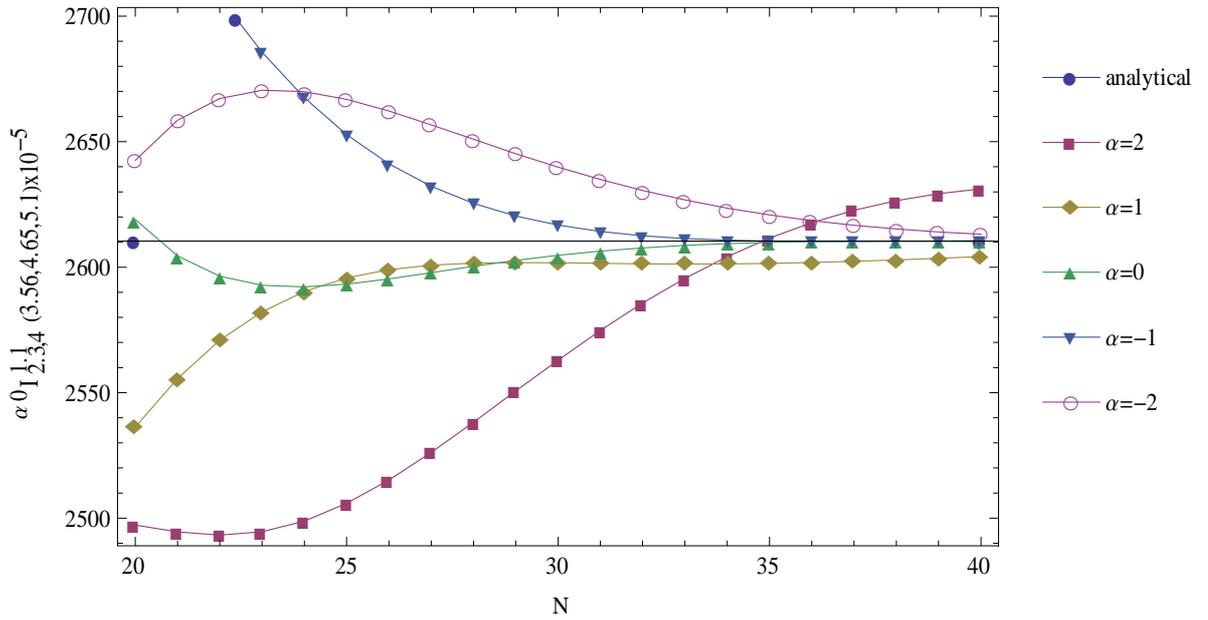

Fig.2. Convergence of arrange (42a) and rearranged (42b) power series for Yukawa like integrals ${}^{\alpha 0}I_{2.3,4}^{1.1}(3.56,4.65,5.1)$ for $-2 \leq \alpha \leq 2$